\documentclass[a4paper,11pt,twoside]{article}
\usepackage{amsmath}
\usepackage{graphicx}
\usepackage{amsbsy}
\usepackage{cite}
\usepackage{amsfonts}
\usepackage{color}
\usepackage{amssymb}
\usepackage{subfigure}
\usepackage{slashed}
\newcommand{\te}[1]{\mathrm{#1}}
\numberwithin{equation}{section}

\begin{document}
\thispagestyle{empty}  
\setcounter{page}{0}  
\begin{flushright}
CP3-12-29\\
\end{flushright}

\vskip 2.2 true cm

\begin{center}
{\huge Weakly-induced strong CP violation}\\[1.9cm]

\textsc{Jean-Marc G\'erard}$^{1}$\textsc{ and Philippe Mertens}$^{2}$%
\\[12pt]\textsl{Centre for Cosmology, Particle Physics and Phenomenology
(CP3), }

\textsl{Universit\'{e} catholique de Louvain, Chemin du Cyclotron 2, \\
 1348 Louvain-la-Neuve, BELGIUM}\\[60pt]

\textbf{Abstract}
\end{center}
\begin{quote}
\noindent \\
Weak interaction contributions to the strong $\theta$ parameter are revisited in the frame of a large-$N_c$ Chiral Perturbation Theory. Focusing on the hadronic $\eta^{(\prime)}\rightarrow \pi \pi$ amplitudes, we express these second-order corrections in terms of the CP-violating parameter in $K \rightarrow \pi \pi$ decays to obtain $\Delta_w \theta \approx 10^{-17}$ at $\mathcal{O}(G_F^2 \varepsilon^\prime)$.

\footnotetext[1]{jean-marc.gerard@uclouvain.be} \footnotetext[2]{philippe.mertens@uclouvain.be}

\end{quote}

\begin{quote}
\noindent\ 

\newpage%
\setcounter{tocdepth}{2}
\end{quote}

\section{Introduction}
Nowadays, the fundamental interactions between elementary particles appear to emerge from a universal gauge principle in a Quantum Field Theory. Within such a theoretical frame, the fascinating microscopic irreversibility due to time-reversal (T) non-invariance or, equivalently, CP violation tends to be associated with short range weak and strong nuclear forces through the quite subtle mechanisms of spontaneous symmetry breaking and confinement, respectively. However, this paradigm may dramatically change soon, with the discovery of an elementary spin zero particle, the so-called Higgs boson. Indeed, its Yukawa couplings with the matter fields are in fact the primary sources of CP violation but do not correspond to a gauge interaction, at least as they stand. In this context, it is therefore worth revisiting a launched bridge between weak and strong CP violation.

In the Standard Model (SM), both the electroweak gauge interactions and the Higgs self interactions turn out to be CP-invariant. Yet, in absence of any flavour theory, the most general Yukawa interactions of the Higgs field with three generations of quarks are responsible for two independent CP-violating phases. The first one, $e^{i\delta_\te{CKM}}$, preserves parity (P) while the second one, $e^{i\theta_\te{QFD}}$, preserves charge conjugation (C). Indeed, these phases are rooted in the complex up (and down) quark mass matrices $M^{u(d)}$ : induced by the Higgs field frozen at its vacuum expectation value, these matrices can always be polar decomposed into Hermitian ones times a global phase \cite{jmgHerbeumont}, but are neither symmetric nor Hermitian.

As a matter of fact, the $\delta_\te{CKM}$ and $\theta_\te{QFD}$ angles are not observables by themselves. On the one hand, the unitarity of the three-by-three Cabibbo-Kobayashi-Maskawa (CKM) mixing matrix allows nine independent parametrizations in terms of Euler rotations such that flavour physics only implies the lower bound \cite{xingjmg}
\begin{equation}
\label{CKM}
 \delta_\te{CKM} \gtrsim \pi/200\;.
\end{equation}
On the other hand, the axial anomaly in strong gauge interactions is such that nuclear physics only requires the upper bound \cite{edm1,edm2}
\begin{equation}
\label{theta}
\theta \equiv \theta_\te{QFD} + \theta_\te{QCD} \lesssim 10^{-10}
\end{equation}
with $\theta_\te{QFD}$, the argument of $\det (M^u M^d)$ in Quantum Flavour Dynamics (QFD) and 
$\theta_\te{QCD}$, the coefficient in front of the $G_{\mu \nu}\Tilde{G}^{\mu \nu}$ term in Quantum Chromo Dynamics (QCD) \cite{thooft}. 

The striking hierarchy between Eq.(\ref{CKM}) and Eq.(\ref{theta}) suggests that $\delta_\te{CKM}\neq 0$ and $\theta=0 $ at the classical level. A natural way to implement such a scenario would be to impose the parity invariance on the full Lagrangian. However, in the SM, C and P discrete symmetries are explicitly broken by the gauge sector such that quantum corrections to the $\theta$ parameter are expected to arise at the second-order in the electroweak interactions. 
In the past, two complementary short-distance attempts to estimate $\Delta_w \theta$ within the SM have been proposed. The first one \cite{ellis} was based on loop corrections for the light quark masses, leading to 
\begin{equation}
\label{ellis}
\Delta_w \theta_\te{QFD} \approx 10^{-16}~~\te{at}~~\mathcal{O}(G_F^2 \alpha_s ^3)\;,
\end{equation}
while the second one \cite{khriplo} has considered the induced gluon pseudo-strength field to get 
\begin{equation}
\label{khriplo}
\Delta_w \theta_\te{QCD}\approx 10^{-19}~~\te{at}~~\mathcal{O}(G_F^2 \alpha_s)\;.
\end{equation} 
In this Letter we estimate $\Delta_w \theta$ through the physical $\eta^{(\prime)}\rightarrow \pi \pi$ hadronic decays and find rather
\begin{equation}
\label{result}
\Delta_w \theta \approx 10^{-17}~~\te{at}~~\mathcal{O}(G_F^2 \varepsilon^\prime)
\end{equation}
with $\varepsilon^\prime$, the penguin-induced CP violation parameter in $K \rightarrow \pi \pi$ decays.
\section{$\eta^{(\prime)}\rightarrow \pi \pi$ from strong interactions}
At low energy, all the basic aspects of strong interactions are encapsulated in the truncated $\mathcal{O}(p^2)$ effective Lagrangian \cite{trahern,veneziano,witten}
\begin{equation}
\mathcal{L}_{S}=\frac{F^2}{4}  \langle \partial_\mu U \partial^\mu U^\dagger \rangle +\mathcal{L}_{S}^M+\mathcal{L}_{S}^{\theta} \;,
\label{stronglag}
\end{equation}
where $\langle A \rangle$ represents the trace of $A$. The unitary field $U$ transforms as a $(3_L,\bar{3}_R)$ multiplet of the chiral $U(3)_L \otimes U(3)_R$ group and its $U(3)_V$-invariant vacuum expectation value (i.e., the unity matrix $\mathbb{I}$) is perturbed by the full nonet $\phi$ of Goldstone bosons (GB) :
\begin{equation}
U=\mathbb{I}+i(\phi/F)-\frac{1}{2}(\phi/F)^2+i b(\phi/F)^3-(b+1/8)(\phi/F)^4+\mathcal{O}\left(\phi^5\right)\;, \label{phi}
\end{equation}
with 
\begin{equation}
\label{9GB}
\phi = \sum_{a=0}^{8}\lambda_a \phi_a =\left(\begin{array}{ccc}
\pi^0+\frac{\eta_8}{\sqrt{3}}+\sqrt{\frac{2}{3}}\eta_0 & \sqrt{2}\pi^+ &\sqrt{2} K^+\\
\sqrt{2}\pi^- &-\pi^0 +\frac{\eta_8}{\sqrt{3}}+\sqrt{\frac{2}{3}}\eta_0  &\sqrt{2} K^0\\
\sqrt{2}K^- &\sqrt{2}~ \overline{K^0} &-\frac{2}{\sqrt{3}}\eta_8 +\sqrt{\frac{2}{3}}\eta_0 
\end{array}\right)\;.
\end{equation}
In our conventions, $\lambda_a$ ($a=1,\dots,8$) are the standard Gell-Mann matrices complemented by $\lambda_0=\sqrt{2/3}~\mathbb{I}$, all the GB have a canonical kinetic term, the constant $F$ is the pion decay constant ($F=F_\pi=92.4~\te{MeV}$) and any physical process must be $b$-independent \cite{coleman1,coleman2}.

\subsection{Mass spectrum and mixing from $\mathcal{L}_{S}^M$}
All nine GB in Eq.(\ref{9GB}) acquire a mass through the following chiral symmetry-breaking terms
\begin{equation}
\label{masslag}
\mathcal{L}_{S}^M=\frac{F^2}{4} \left[\langle \mu^2 ( U +  U^\dagger)\rangle + \frac{m_0^2}{4 N_c} \langle \ln{U}-\ln{U^\dagger}\rangle^2\right] \;.
\end{equation}
On the one hand, the vacuum expectation value of the $\mu^2$ matrix field is proportional to the real and diagonal light quark mass matrix and provides the pions and kaons with a mass :
\begin{equation}
\mu^2_u=\mu^2_d=m_\pi^2~~\te{and}~~\mu^2_s=2m_K^2-m_\pi^2\;. \label{Kpimasses}
\end{equation}
As such, it breaks the flavour $SU(3)$ symmetry but preserves its isospin sub-group $SU(2)_I$ in the limit  
$\mu^2_u = \mu^2_d$. On the other hand, the colour-suppressed operator proportional to $m_0^2$ in Eq.(\ref{masslag}) is responsible for the breaking of the anomalous axial $U(1)_A$ subgroup of $U(3)_L \otimes U(3)_R$ and allows us to consider $\eta_0$ as the ninth GB of the $U(3)$ multiplet $\phi$ in the large-$N_c$ limit \cite{wittenNc}. However, since $\eta_8$ and $\eta_0$ mix, it is suitable to introduce the single mixing angle $\varphi$ which relates the $SU(3)$ eigenstates $(\eta_8, \eta_0)$ and the mass eigenstates $(\eta, \eta^\prime)$ in the isospin limit as
\begin{equation}
\left(\begin{array}{c}
\eta \\ \eta^\prime
\end{array} \right) =\left( \begin{array}{cc}
\cos \varphi & - \sin \varphi \\
\sin\varphi & \cos\varphi
\end{array}\right) \left( \begin{array}{c}
\eta_8 \\ \eta_0
\end{array}\right)  \;~~~~\te{with}~~~~-\frac{\pi}{4} <\varphi <\frac{\pi}{4}\;.
\label{eigen}
\end{equation}  
We then obtain the following $\mathcal{O}(p^2)$ mass spectrum for the iso-singlet states
\begin{subequations}
\label{etaetapmass}%
\begin{align}
m_{\eta^\prime}^2&=\frac{1}{3}\left(4m_K^2-m_\pi^2-2 \sqrt{2}(m_K^2-m_\pi^2)\cot \varphi\right) \label{metap}\\
m_\eta^2 &=\frac{1}{3}\left(4m_K^2-m_\pi^2+2 \sqrt{2}(m_K^2-m_\pi^2)\tan \varphi\right)\;, \label{meta}
\end{align}
\end{subequations}
with the mixing angle $\varphi$ and the scale parameter $m_0$ intimately related through
\begin{equation}
\tan 2 \varphi = 2 \sqrt{2}\left[ 1-\frac{3}{2}\frac{3}{N_c}\frac{m_0^2}{m_K^2-m_\pi^2}\right]^{-1}\;.
\end{equation}
Interestingly, the Eqs.(\ref{etaetapmass}) allow for two mass degeneracies :
\begin{equation}
\label{mixingangle}
\begin{array}{llllll}
1) & m_{\eta^\prime} = m_\pi & \mathrm{when} & m_0^2=0& \mathrm{or}& \varphi=+35.3^\circ\;, \\[8pt]
2) & m_{\eta       } = m_K  & \mathrm{when}  & m_0^2=3(m_K^2-m_\pi^2)& \mathrm{or}& \varphi=-19.5^\circ\;.
\end{array}
\end{equation}
The first one ($m_{\eta^\prime}=m_\pi$) at the source of the so-called $U(1)_A$ problem \cite{weinberg} requires a non-vanishing $m_0$ parameter. More precisely, in order to reproduce the $\eta^\prime(958)$ mass we should set $m_0=817~\te{MeV}$ or equivalently $\varphi\approx -20^\circ$ if the physical masses for $K(498)$ and $\pi(135)$ are imposed in Eq.(\ref{metap}). This particular mixing angle turns out to be very close to the one at which the second degeneracy ($m_{\eta} = m_K$) occurs. However, $m_\eta^2$ almost fulfills the Gell-Mann-Okubo (GMO) mass relation $m_{88}^2=(4 m_K^2-m_\pi^2)/3$. So, to reproduce exactly the mass of $\eta(548)$ one should rather impose a mixing angle close to zero in Eq.(\ref{meta}). In other words, the physical mass spectrum for $\eta(548)$, $\eta^\prime(958)$, $K(498)$ and $\pi(135)$ cannot be simultaneously reproduced within the truncated frame adopted here. This can be nicely quantified by the $\varphi$-independent upper bound \cite{georgi,jmgkou} 
\begin{equation}
\frac{m_\eta^2-m_\pi^2}{m_{\eta^\prime}^2-m_\pi^2}< 2-\sqrt{3}\simeq 0.27
\end{equation}
which calls for a $20\%$ correction to be compatible with the measured ratio 0.33. 
To accommodate the full nonet mass spectrum, higher-order operators such as
\begin{equation}
\langle \mu^2 (U-U^\dagger)\rangle  \langle\ln{U}-\ln{U^\dagger} \rangle \ni \langle \phi \rangle ~\eta_0  \label{su3breaking1}
\end{equation}
have to be considered \cite{jmgkou}. Yet, this $\mathcal{O}(p^2,1/N_c)$ operator together with $\mathcal{O}(p^4,0)$ ones will not be considered in this Letter since the effective Lagrangian in Eq.(\ref{stronglag}) is restricted to the leading $\mathcal{O}(p^2,0)$  and $\mathcal{O}(p^0,1/N_c)$  terms, respectively.
\subsection{CP-violating interactions from $\mathcal{L}_{S}^{\theta}$}
Once we ensure that none of the GB acquire a vacuum expectation value thanks to a suited $U$ field phase redefinition, the full effect of the strong $\theta$ angle can be encoded into \cite{pichcp}
\begin{equation}
\mathcal{L}_{S}^{\theta}= i K_\theta \frac{F^2}{4}  \left[-\langle U-U^\dagger\rangle + \langle \ln{U}-\ln{U^\dagger}\rangle\right]
\end{equation}
which contains no linear term in $\phi$. At this level, any P- and T-violating observable quantity will thus depend on the constant factor $K_\theta$ rather than on the $\theta$ parameter itself.
Consequently, in the SM the first and simplest manifestation of a non-zero $\theta$ is the occurrence of C-conserving two-body decays via the strong interaction term 
\begin{equation}
\mathcal{L}_{S}\left|_{\phi^3} \right.=\mathcal{L}_{S}^{\theta}\left|_{\phi^3} \right. = - \frac{K_\theta}{3\sqrt{2}  F} \langle \phi^3 \rangle 
\end{equation}
that does not depend on the b-parametrization chosen for the $U$ matrix expanded in Eq.(\ref{phi}). Solely $\eta^{(\prime)} \rightarrow \pi \pi$ on-shell decays are allowed by energy conservation and we obtain the pure $\Delta I =0$ strong amplitudes
\begin{subequations}
\label{strongampli}%
\begin{equation}
A(\eta^\prime \rightarrow \pi \pi)_s =  \frac{K_\theta}{\sqrt{3}F} \left(s_\varphi + \sqrt{2}~c_\varphi \right)
\end{equation}
\begin{equation}
A(\eta \rightarrow \pi \pi)_s =  \frac{K_\theta}{\sqrt{3}F} \left(c_\varphi - \sqrt{2}~s_\varphi \right)\;,
\end{equation}
\end{subequations}
having set $s_\varphi=\sin \varphi$ and $c_\varphi=\cos \varphi$ for short.
By comparing the subsequent prediction $\te{\Gamma}(\eta \rightarrow \pi^+ \pi^-)\simeq 2.6 ~|K_\theta|^2 ~\te{GeV}^{-3}$, obtained for the phenomenological mixing angle $\varphi \approx -20^\circ$, with the experimental limit $\te{Br}(\eta \rightarrow \pi^+ \pi^-) < 1.3 \times 10^{-5}$ \cite{PDG} we infer the upper bound $K_\theta \lesssim 2.6 \times 10^{-6}~ \te{GeV}^2$. As a consequence, $K_\theta$ is small enough to be approximated by \cite{pichcp,jmgHerbeumont}
\begin{equation}
K_\theta=\frac{m_\pi^2}{2}~\theta \;,
\end{equation}
in the realistic limit $\mu_u^2 = \mu_d^2 \ll \mu_s^2,m_0^2$.

The $(\eta,\eta^\prime)$ mass eigenstates being complementary in the trigonometric sense, see Eq.(\ref{eigen}), we conclude that the relation
\begin{equation}
A(\eta \rightarrow \pi \pi)=\left. A(\eta^\prime \rightarrow \pi \pi) \right|_{\varphi \rightarrow \varphi + \frac{\pi}{2} } \;, \label{frometaptoeta}
\end{equation}
fulfilled by Eqs.(\ref{strongampli}), constitutes a good cross-check for our forthcoming computations.
Note also that the mixing angle dependences appearing in Eqs.(\ref{strongampli}) are specific to the single anomalous term ($\mathcal{L}_{S}^{\theta} \ni \langle \phi^3 \rangle$) appearing at order $\mathcal{O}(p^0,1/N_c)$. In principle, other mixing angle dependences can be induced. For example, the P- and T-violating operator going along with the $\mathcal{O}(p^2,1/N_c)$ one in Eq.(\ref{su3breaking1}), namely
\begin{equation}
\langle  U+U^\dagger-2\rangle \langle\ln{U}-\ln{U^\dagger} \rangle \ni \langle \phi^2 \rangle ~\eta_0 \;,
\label{su3break}
\end{equation}
generates pure $\cos \varphi$ ($\sin \varphi$) contribution to $A(\eta^{(\prime)} \rightarrow \pi \pi)_s$. This observation will be of some relevance in our confrontation with the weak interaction contributions to these decay processes.

\section{$\eta^{(\prime)}\rightarrow \pi \pi$ from weak interactions}
Such CP-violating but flavour-conserving weak processes require a two step change of flavour \cite{shaba}.
At low energy, the $|\Delta S|=1$ weak interactions involving the GB are also ruled by the chiral $U(3)_L \otimes U(3)_R$ transformations acting on the $U$ field. These interactions are encoded in the $\mathcal{O}(p^2)$ effective Hamiltonian given by \cite{jmgcs}
\begin{equation}
\mathcal{H}_{W}^{|\Delta S| =1} = g_8 Q_8+ g_{27}Q_{27} + g_s Q_s  +g_\mu Q_\mu + \te{h.c.} \label{WeakLagrangian}
\end{equation}
where the standard operators 
\begin{subequations}%
\begin{align}
Q_8 &= (L_\alpha L^\alpha)_{23} \\
Q_{27}&= ({L_\alpha})_{23} (L^\alpha)_{11} +\frac{2}{3}(L_\alpha)_{13} (L^\alpha)_{21}-\frac{1}{3}(L_\alpha)_{23} \langle L^\alpha \rangle
\end{align}
\end{subequations}
are built up from the left-handed hadronic currents normalized to $L_\alpha \equiv i F^2 \partial_\alpha U U^\dagger$.
Besides these operators known to saturate the $K\rightarrow \pi \pi$ decay amplitudes in the isospin limit, the current-current operator 
\begin{equation}
Q_s = (L_\alpha)_{23} \langle L^\alpha \rangle
\end{equation}
is proportional to the flavour singlet $\eta_0$ field, while the mass operator 
\begin{equation}
Q_\mu= F^4(\mu^2 U^\dagger +U \mu^2)_{23}
\end{equation}
does not contribute to $\mathcal{O}(G_F)$ on-shell amplitudes \cite{crewther}.
\begin{figure}[t]
\centering
\subfigure[]{\includegraphics[scale=0.3]{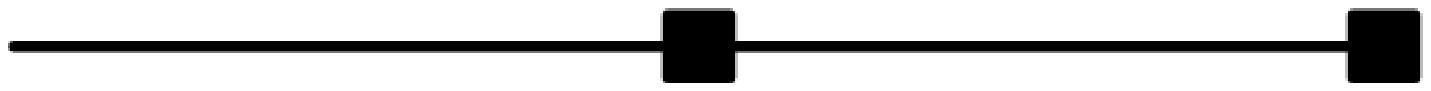}}~~~~~~~~~~~~
\subfigure[]{\includegraphics[scale=0.3]{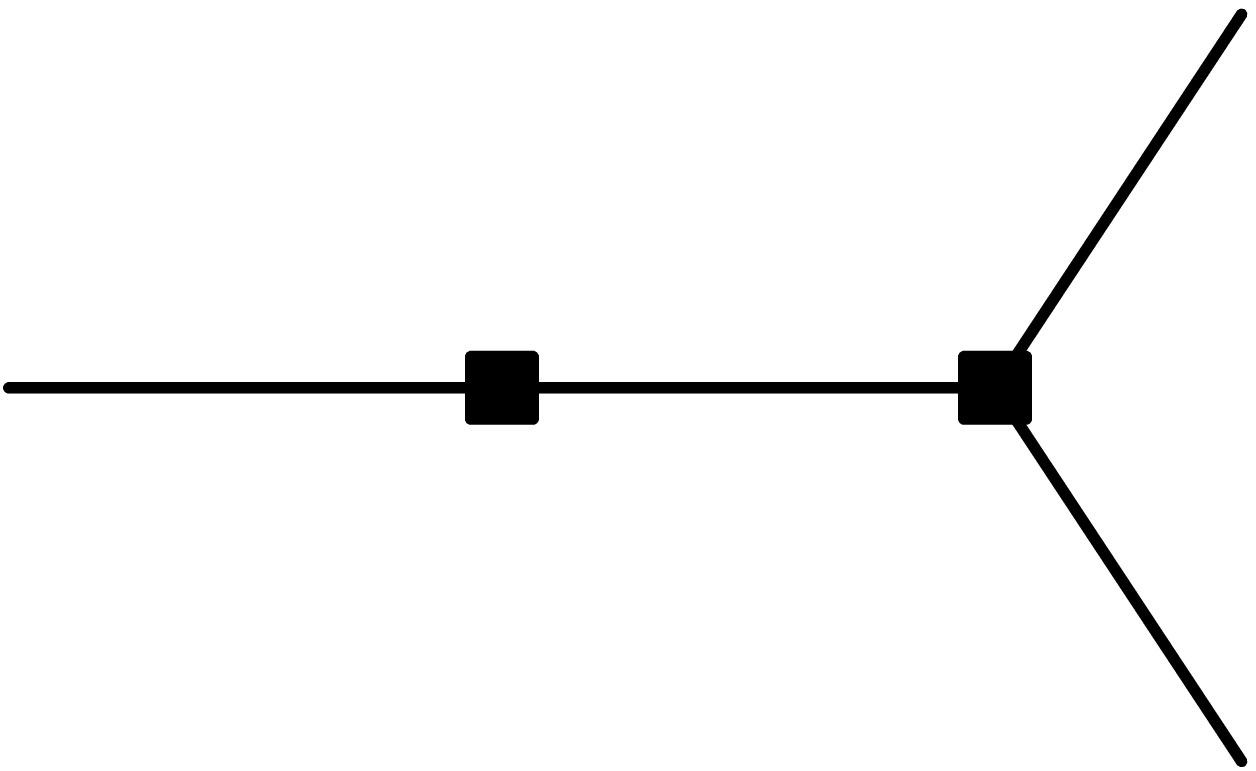}} ~~~~~~~~~~~~
\subfigure[]{\includegraphics[scale=0.3]{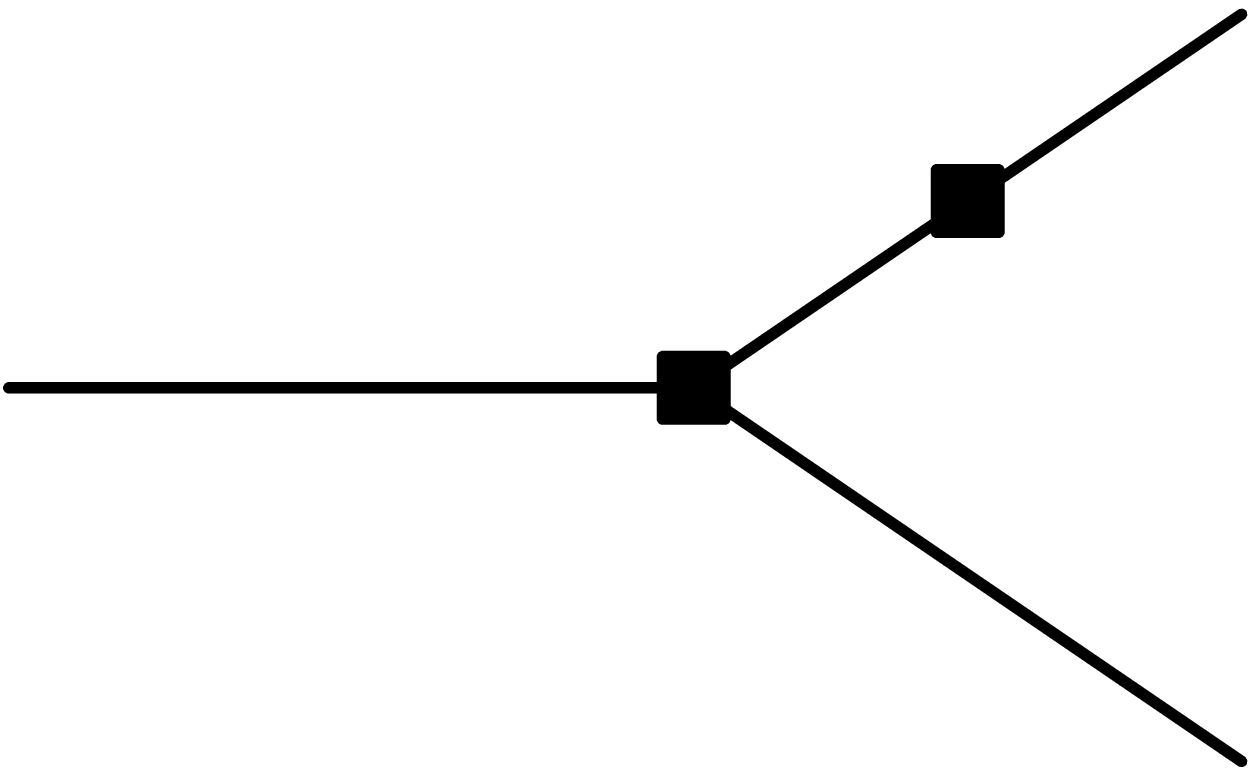}} \\
\subfigure[]{\includegraphics[scale=0.3]{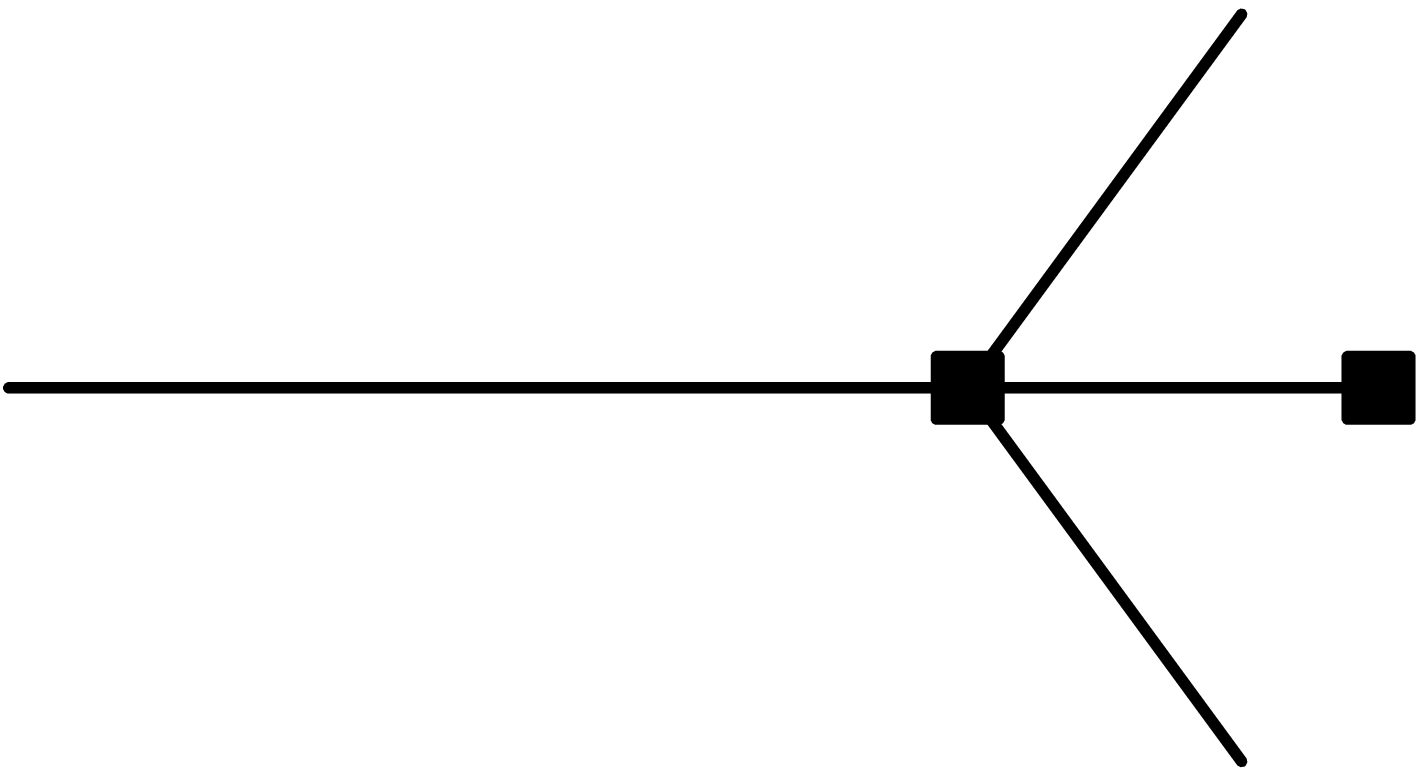}} ~~~~~~~~~~~~
\subfigure[]{\includegraphics[scale=0.3]{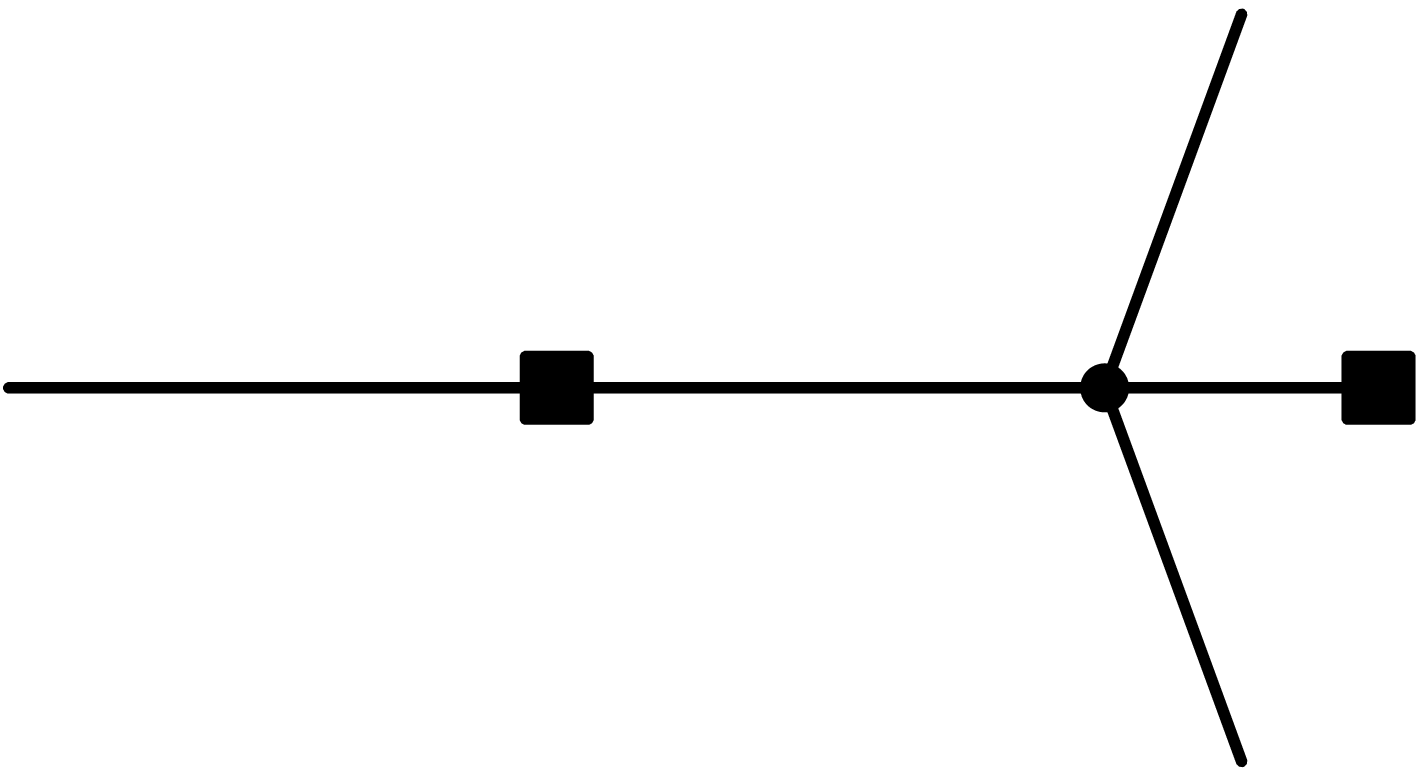}} ~~~~~~~~~~~~
\subfigure[]{\includegraphics[scale=0.3]{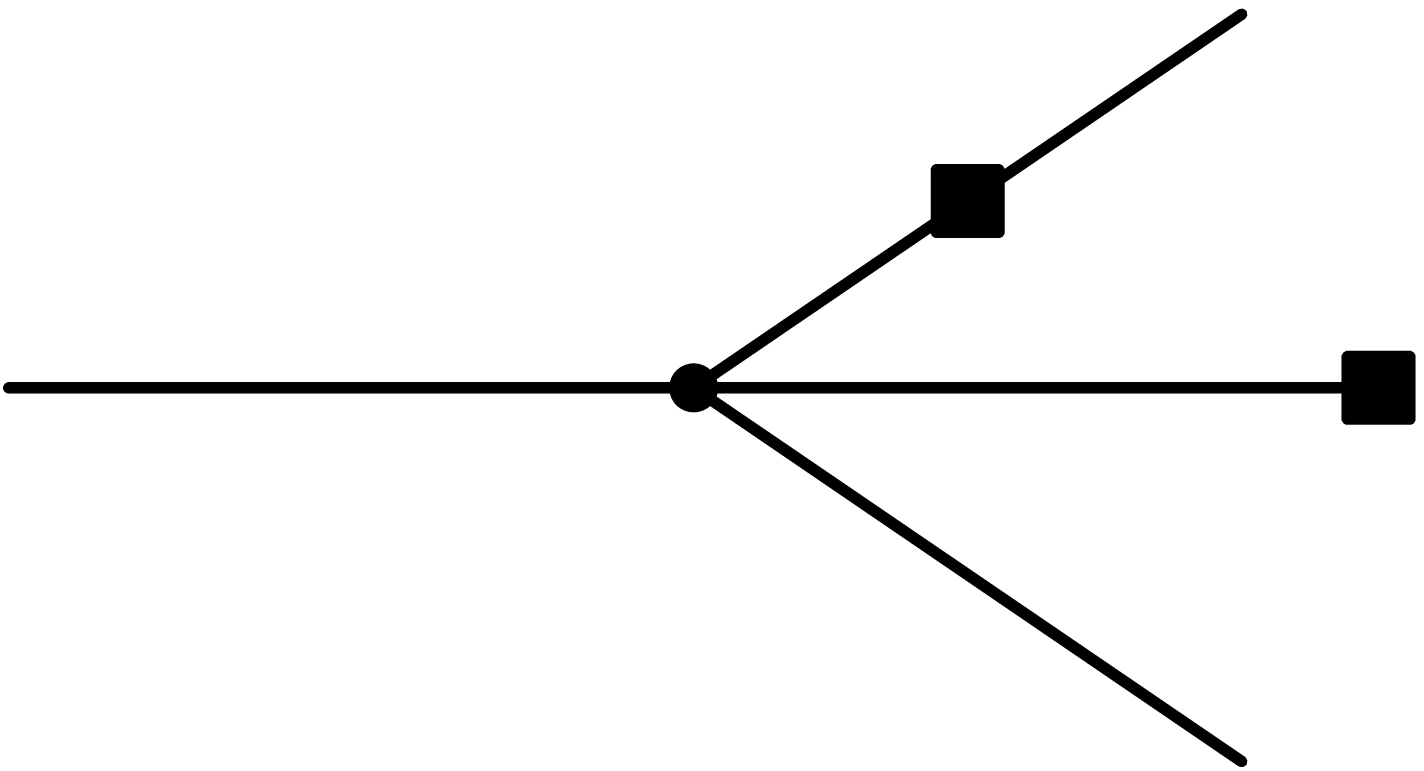}} 
\caption{The $\mathcal{O}(p^2)$ topologies generating (a) possible tadpole amplitudes and (b-f) two-body decay amplitudes. Black squares stand for weak vertices induced by $\mathcal{H}_{W}^{|\Delta S|=1}$ in Eq.(\ref{WeakLagrangian}), whereas black dots represent strong vertices induced by $\mathcal{L}_S$ in Eq.(\ref{stronglag}).}
\label{fg}
\end{figure}

In order to generate $\Delta S =0$ and CP-violating amplitudes from the Hamiltonian given in Eq.(\ref{WeakLagrangian}), successive $\Delta S =\pm 1$ and $\Delta S =\mp 1$ transitions must interfere in such a way that 
\begin{equation}
A(i\rightarrow f)_w=\sum_{I \neq J} A_{IJ}(i\rightarrow f)~ \te{Im}(g_I^* g_J)\;. \label{am}
\end{equation}
As a first consequence, the isospin-breaking electroweak operator
\begin{equation}
Q_{ew}=F^6 \alpha_\te{e.m.} (U^\dagger Q U)_{23}\;,\label{elec}
\end{equation}
with $Q=(2/3,-1/3,-1/3)$, can be neglected as far as the $\eta^{(\prime)}\rightarrow \pi \pi $ decays are concerned. Indeed, this operator does not affect the neutral $\eta^{(\prime)}\rightarrow \pi^0 \pi^0 $ decay amplitudes and only decreases the charged ones by less then ten percent. The main reason is that, contrary to the CP-violating parameter $\varepsilon^\prime$ proportional to the ratio of $g_I$ effective couplings, the CP-violating $\eta^{(\prime)}\rightarrow \pi \pi $ decay amplitudes are proportional to their product. So, here there is no possible $\varepsilon^\prime$-like $\Delta I =1/2$ enhancement to compensate for the naive $\alpha_\te{e.m.}/\alpha_s$ suppression factor.\par
Among the second-order weak amplitudes in Eq.(\ref{am}), tadpole-like ones shown in Fig.\ref{fg}a vanish trivially. Indeed, any inclusion of a current-current $Q_I$ operator ($I \neq \mu$) in the $\pi^0,\eta^{(\prime)} \rightarrow K$ vertex generates an amplitude proportional to the square of the incoming four-momentum.

Considering now the two-body decays generated by the non-local topologies displayed in Fig.\ref{fg}b-f, we obtain at $\mathcal{O}(p^2)$ and in the isospin limit the tree-level weak $\eta^\prime$ amplitudes
\begin{subequations}
\label{weakampli}
\begin{align}
A(\eta^{\prime} \rightarrow \pi^+ \pi^-)_w &= \frac{4}{3\sqrt{3}}F^3 \alpha(m_{\eta^\prime}^2)\left[5 I_{8,27}~s_\varphi-\left(4I_{8,27}-9 I_{8,s}-6I_{27,s} \right)\sqrt{2} ~c_\varphi \right]\\
A(\eta^{\prime} \rightarrow \pi^0 \pi^0)_w  &= \frac{4}{3\sqrt{3}}F^3 \alpha(m_{\eta^\prime}^2)\left[6 I_{8,27}+9 (I_{8,s}-I_{27,s})\right]~\sqrt{2}~c_\varphi
\end{align}
\end{subequations}
where
\begin{equation}
\label{alpha}
\alpha(p^2)\equiv p^2 \left(\frac{p^2-m_\pi^2}{p^2-m_K^2}\right)
\end{equation}
and
\begin{equation}
\label{IIJ}
I_{I,J}\equiv \te{Im}(g_I^* g_J)\;.
\end{equation}
As explicitly checked, the weak $\eta \rightarrow \pi \pi$ amplitudes fulfill the complementary relation mentioned in Eq.(\ref{frometaptoeta}), namely they are deduced from Eqs.(\ref{weakampli}) after replacing $s_\varphi$ by $c_\varphi$ and $c_\varphi$ by $-s_\varphi$.

In Eq.(\ref{alpha}), the simple pole at the K-mass indicates that the only relevant topology for $\eta^{(\prime)}\rightarrow \pi \pi$ decays is eventually the one depicted on Fig.\ref{fg}b. Indeed, the remaining topologies of Fig.\ref{fg}c-f cancel out  once summed up, making the $\mathcal{O}(G_F^2)$ amplitudes given in Eqs.(\ref{weakampli}) $b$-independent, as it should be, but also $g_\mu$-independent.

\section{From $\delta_\te{CKM}$ to $\theta_{QCD}$}
As proved in the previous section, weak interactions do contribute to the P- and T- violating $\eta^{(\prime)}\rightarrow \pi \pi$ decays at second-order. Therefore, within the SM, these weak  corrections contribute to the $K_\theta$ parameter or, equivalently, to the strong $\theta$ term. In this section, we show how this can be achieved assuming again both the isospin and large-$N_c$ limits.

To begin with, let us have a first look at the $g_{I}$ effective coupling constants. Below the charm mass scale, the QCD-induced $|\Delta S| =1$ effective Hamiltonian approximatively reads
\begin{eqnarray}
\label{effectiveham2}
\mathcal{H}_W^{|\Delta S|=1}(\mu < m_c)  &\simeq &\frac{G_F}{\sqrt{2}}~~ \{V_{ud}V_{us}^\ast  \left[ z_1 (\mu) Q_1(\mu) +z_2(\mu) Q_2(\mu)\right]  + \nonumber  \\ 
&& ~~~~~+ \left[V_{ud}V_{us}^\ast~z_6(\mu)-V_{td}V_{ts}^\ast~y_6(\mu)  \right] Q_6(\mu)\}+\te{h.c.}\;
\end{eqnarray}
where $V$ is the unitary CKM mixing matrix. The Wilson coefficients associated with the current-current $Q_{1,2}$ \cite{gaillardlee,altarellimaiani} and density-density $Q_6$ \cite{vainshtein} four-quark operators are denoted by $z_{1,2}$ and $z_6,y_6$, respectively. They encode the short-distance (SD) evolution from $M_W$ down to a $\mu$ scale above one $\te{GeV}$. A comparison with the effective Hamiltonian given in Eq.(\ref{WeakLagrangian}) should in principle allow us to assign the CKM phase to the $g_I$ couplings. However, to do so, we first have to include the long-distance (LD) evolution down to the hadronization scale 
$\mu_\te{had}$ lying well below one $\te{GeV}$ where perturbative QCD breaks down. Fortunately, Chiral Perturbation Theory supplemented with the $1/N_c$ expansion allows us to go from the quark-gluon picture to the meson one to get \cite{jmgmeson,acta}
\begin{equation}
\label{had}
\mathcal{H}^{|\Delta S|=1}(\mu_\te{had})\simeq \frac{G_F}{\sqrt{2}}~~ \{ x_1 \hat{Q}_1 +x_2 \hat{Q}_2 +x_6 \hat{Q}_6 \}+\te{h.c.}\;,
\end{equation}
with
\begin{equation}
\hat{Q}_1=({L_\alpha})_{23} (L^\alpha)_{11},\; \hat{Q}_2=({L_\alpha})_{13} (L^\alpha)_{21},\;\hat{Q}_6=({L_\alpha}L^\alpha)_{23}\;.
\end{equation}
In other words, no additional chiral structures appear beyond the one already present in Eq.(\ref{WeakLagrangian}) since the product of quark currents factorizes into a product of meson ones at the hadronization scale. At this scale, we thus have a one-to-one formal correspondence between the $g_I$ effective couplings and the $x_i$ (SD plus LD) coefficients \cite{jmgcs} :
\begin{equation}
g_8\simeq \frac{G_F}{\sqrt{2}}\left[ -\frac{2}{5}x_1 + \frac{3}{5}x_2+x_6\right]\;,~~~g_{27}\simeq  \frac{G_F}{\sqrt{2}} \left[\frac{3}{5}(x_1+x_2)\right]\;,~~~g_s \simeq \frac{G_F}{\sqrt{2}} \left[\frac{3}{5}x_1 -\frac{2}{5}x_2 \right] 
\;.\label{wilcof}
\end{equation}
Within this matching approach, the CP-violating $I_{I,J}$ elements defined in Eq.(\ref{IIJ}) arise then exclusively from quark $Q_{1,2}-Q_6$ interference (i.e., $ V_{ud}V_{us}^\ast- V_{td}V_{ts}^\ast$ interference) such that the subdominant $27-s$ meson topology of Fig.1b is real :
\begin{equation}
\label{i27s}
I_{27,s}=0 \;.
\end{equation} 

Let us now use the strong amplitudes given in Eq.(\ref{strongampli}) as a guideline to identify the weak contributions to the $K_\theta$ parameter. The exhibited isospin symmetry between charge and neutral pion final states might be enforced on the $\eta \rightarrow \pi \pi$ amplitudes :
\begin{equation}
A(\eta \rightarrow \pi^+ \pi^-)_w = A(\eta\rightarrow \pi^0 \pi^0)_w =\frac{4}{3}\sqrt{\frac{2}{3}}F^3 \alpha(m_{\eta}^2)\left[2 I_{8,27}+3 I_{8,s} \right]\;,
\end{equation}
if and only if we assess a specific mixing angle, i.e., 
\begin{equation}
\tan \varphi = -\frac{1}{2 \sqrt{2}}~~\mathrm{or}~~\varphi=-19.5^\circ\;,
\end{equation}
as been done in \cite{shaba}.
However, this phenomenological angle is rather problematic here since the $A(\eta \rightarrow \pi \pi)_w$ amplitude proportional to $\te{Im}(x_6^\ast x_1)$ would then develop a pole (see Eqs.(\ref{mixingangle}) and (\ref{alpha})). Moreover, it would also imply  $A(\eta^\prime \rightarrow \pi^+ \pi^-)_w \neq A(\eta^\prime \rightarrow \pi^0 \pi^0)_w$. So, we find more appropriate to isolate the $\Delta I =0$ component of the $\eta^{(\prime)}\rightarrow \pi \pi$ weak amplitudes :
\begin{subequations}
\label{iso0}
\begin{align}
A(\eta^\prime \rightarrow \pi \pi)_{w}^0 &= \frac{4}{3\sqrt{3}}F^3 \alpha(m_{\eta^\prime}^2)\left[\frac{10}{3} I_{8,27}(s_\varphi+\sqrt{2}~c_\varphi)-\left(4 I_{8,27}-9I_{8,s} \right)~\sqrt{2}~c_\varphi \right] \\
A(\eta\rightarrow \pi \pi)_{w}^0 &= \frac{4}{3\sqrt{3}}F^3 \alpha(m_{\eta}^2)\left[\frac{10}{3} I_{8,27}(c_\varphi-\sqrt{2}~s_\varphi)+\left(4 I_{8,27}-9I_{8,s} \right)~\sqrt{2}~s_\varphi \right]\;,
\end{align}
\end{subequations}
leaving aside their $\Delta I =2$ components explicitly given by 
\begin{equation}
A(\eta\rightarrow \pi^+ \pi^-)_{w}^2 = - 2~ A(\eta\rightarrow \pi^0 \pi^0)_{w}^2= \frac{20}{9\sqrt{3}}F^3 \alpha(m_{\eta}^2)~I_{8,27}~(c_\varphi+2\sqrt{2}~ s_\varphi)
\end{equation}
in the case of $\eta \rightarrow \pi \pi$.
A direct identification based now on the mixing angle dependence of Eqs.(\ref{strongampli}) provides then the $\mathcal{O}(G_F^2)$ corrections to the strong P- and T- violating amplitudes : 
\begin{subequations}
\label{tot}
\begin{align}
\label{cometap}
\Delta_w A(\eta^\prime \rightarrow \pi \pi)&= \frac{40}{9\sqrt{3}}F^3~ I_{8,27}~\alpha(m_{\eta^\prime}^2) ~(s_\varphi+\sqrt{2}~c_\varphi) \\
\Delta_w A(\eta \rightarrow \pi \pi)&= \frac{40}{9\sqrt{3}}F^3~I_{8,27}~ \alpha(m_{\eta}^2)~ (c_ \varphi-\sqrt{2}~s_\varphi)\;.\label{cometa}
\end{align}
\end{subequations}
Note that the pure $\sin \varphi$ ($\cos \varphi$) component of $A(\eta^{(\prime)} \rightarrow \pi \pi)_w^0$ will affect contributions induced by a strong operator like the one given in Eq.(\ref{su3break}). 

Still, contrary to what is predicted by the strong amplitudes in Eqs.(\ref{strongampli}), the coefficients in front of the mixing angles in Eqs.(\ref{tot}) do not match exactly if the $\eta$ and $\eta^{\prime}$ physical masses are enforced :
\begin{equation}
\alpha(m_\eta^2) = 1.62 ~ \te{GeV}^2~~~~~~\neq ~~~~~~1.23 ~ \te{GeV}^2=\alpha(m_{\eta^\prime}^2)\;.
\end{equation}
In other words, a $30\%$ splitting in the effective $\Delta_w K_\theta$ factor is obtained if the physical mass spectrum for the $\eta(548)$, $\eta^\prime(958)$, $K(498)$ and $\pi(135)$ states is imposed. However, we have already noted that this assumption is not allowed in the truncated theory adopted here. Besides, $\alpha(p^2)$ turns out to be rather unstable against $p^2$ variations around the physical value of $m_\eta^2$. For illustration, allowing the $\eta$ mass to be equal to the GMO prediction, i.e., $570~ \te{MeV}$, we obtain $\alpha(m_{88}^2)=1.30~ \te{GeV}^2$, namely a value closer to $\alpha(m_{\eta^\prime}^2)$.
For these reasons, we use the less sensitive $\Delta_w A(\eta^\prime \rightarrow \pi \pi)$ amplitude to conclude that weak interactions shift $\theta$ by the amount 
\begin{equation}
\Delta_w \theta = \frac{2}{m_\pi^2}\Delta_w K_\theta = \frac{80}{9} \frac{\alpha(m_{\eta^\prime}^2)}{m_\pi^2}~F^4 ~I_{8,27}\;.
\end{equation} 
From the formal correspondence relations given in Eq.(\ref{wilcof}) we have in addition that
\begin{equation}
I_{8,27}\equiv \te{Im}(g_8^\ast g_{27})=\frac{3}{10}G_F^2~\te{Im}\left[x_6^\ast~(x_1+x_2)\right] \;,
\end{equation}
with the $x_{1,2,6}$ coefficients defined at the hadronization scale, namely around $m_{K,\pi}$. So, at this stage, either we exploit information from the SD evolution to infer an upper bound on $\Delta_w \theta$ or we extract these coefficients from the available data to get an estimate of it.  

\subsection{An upper bound on $\Delta_w \theta$}
Let us leave aside LD evolution effects by directly matching the Hamiltonians given in Eqs.(\ref{effectiveham2}) and (\ref{had}). As far as the $Q_6$ penguin operator is concerned, that won't do any harm since the $\mu$ dependence of its Wilson coefficient (almost) cancels the one of the corresponding hadronic matrix element \cite{jmgmeson2}. We then obtain
\begin{equation}
x_6 \simeq - 4\left( \frac{m_K}{1~\te{GeV}}\right)^2\frac{m_K^2}{[m_s+m_d]^2}~ \left[V_{ud} V_{us}^\ast~z_6   -  V_{td}V_{ts}^\ast ~y_6  \right]\;,
\end{equation}
with
\begin{equation}
z_6(1~\te{GeV})\simeq -0.02\;,~ y_6(1~\te{GeV})\simeq -0.10
\end{equation}
obtained by using the naive dimensional reduction scheme \cite{Buchalla}, and 
\begin{equation}
(m_s+ m_d)(1~\te{GeV})\simeq 131 ~\te{MeV}\;,
\end{equation}
by letting the lattice quark masses  given in \cite{flag} evolve down to the $\te{GeV}$ scale.
Regarding the $Q_1+Q_2$ combination, what we know from perturbative QCD is that its Wilson coefficient smoothly decreases as $\mu$ is decreasing (see the $\Delta I=1/2$ rule). Therefore, by imposing 
\begin{equation}
x_1+x_2<(z_1+z_2)(1~\te{GeV})\times V_{ud} V_{us}^\ast \;,
\end{equation} 
where
\begin{equation}
(z_1+z_2)(1~\te{GeV}) \simeq 0.76\;,
\end{equation}
we can infer the upper bound 
\begin{equation}
I_{8,27}  \lesssim  0.32 \times G_F^2 \times J(\delta_\te{CKM}) \;.\label{pert}
\end{equation}
The necessity for the CKM phase to appear only through the Jarlskog invariant \cite{J}
\begin{equation}
J(\delta_\te{CKM})\equiv \te{Im}(V_{ts}^\ast V_{td} V_{ud}^\ast V_{us}) 
\end{equation}
explains, a posteriori, why one has to go to the second-order in the weak interactions to induce a correction to the physical strong $\theta$ parameter. Such would not be the case if other sources of CP violation beyond the SM were considered \cite{ellis}. Taking $J=(2.91^{+0.19}_{-0.11})\times 10^{-5}$ from \cite{PDG}, we then infer the rather conservative bound  
\begin{equation}
\label{upperbound}
\Delta_w \theta < 6 \times 10^{-17}\;.
\end{equation}

\subsection{An estimate of $\Delta_w \theta$}
To this end, let us first extract the $g_{8,27}$ effective couplings from the isospin decomposition of the $K\rightarrow \pi \pi$ decay amplitudes : 
\begin{subequations}
\begin{align}
A(K^0 \rightarrow \pi^+ \pi^-)&=A_0 +  \frac{1}{\sqrt{2}}A_2e^{-i \delta}  \\
A(K^0 \rightarrow \pi^0 \pi^0)&=A_0 - \sqrt{2} A_2e^{-i \delta}\;.
\end{align}
\end{subequations}
Using the $\mathcal{O}(p^2)$ Hamiltonian given in Eq.(\ref{WeakLagrangian}), we obtain
\begin{subequations}
\label{ppp}%
\begin{align}
A_0&=\sqrt{2}F(m_K^2-m_\pi^2)(g_8+g_{27}/9) \\
A_2&=10 F(m_K^2-m_\pi^2)g_{27}/9 \;,
\end{align}
\end{subequations}
such that the measured $K \rightarrow \pi \pi$ decay widths are well reproduced if \cite{ciri}
\begin{equation} 
|g_8|_\te{exp}=0.77~ G_F\;~~~~~~,~~~~~~|g_{27}|_\te{exp}=0.044~ G_F~~~~~~\te{and}~~~~~~\delta_\te{exp} =47.5^\circ \;.
\end{equation}
To go further and extract the imaginary part of $g_{8,27}$ we need to consider the CP-violating observable $\varepsilon^\prime$. It turns out \cite{jmgBur} that $\varepsilon^\prime$ is theoretically well reproduced in the isospin limit provided we compute the hadronic matrix elements in the large-$N_c$ limit, i.e., at the hadronization scale. It is therefore legitimate to expect a rather consistent and reliable estimate for $I_{8,27}$. As a matter of fact, we have at our disposal a CKM convention-independent direct CP-asymmetry, namely
\begin{equation}
\label{bref}
\te{Re}(\varepsilon^\prime)=\frac{1}{\sqrt{2}}~\te{Im}\left(\frac{A_2}{A_0}\right)\sin \delta\;.
\end{equation}
 Plugging now Eqs.(\ref{ppp}) in Eq.(\ref{bref}), we then roughly get 
\begin{equation}
 I_{8,27} \simeq  G_F^2 \times \te{Re}(\varepsilon^\prime)\;,
\end{equation}
in the limit $\te{Im}g_{8} \gg \te{Im}g_{27}$.\\
To be more precise, we have to take into account the fact that the electroweak penguins interfere destructively with the strong one in $\varepsilon^\prime$ \cite{ciri}. Including their leading effect through the operator given in Eq.(\ref{elec}), we can extract $I_{8,27}$ from $\varepsilon^\prime$ :
\begin{equation}
 I_{8,27} = (1.7~\te{to}~2.8)\times G_F^2 \times \te{Re}(\varepsilon^\prime)\;.
\end{equation}
Taking $\te{Re}(\varepsilon^\prime)=(2.5 \pm 0.4)\times 10^{-6}$ from \cite{PDG}, we finally obtain values compatible with the upper bound given in Eq.(\ref{upperbound}), namely
\begin{equation}
\Delta_w \theta =  (2.0~\te{to}~4.6)\times 10^{-17}\;.
\end{equation} 
\section{Conclusion}
The strong CP problem, i.e., the smallness of $\theta$,  is a long-standing one \cite{theta} and scenarios going beyond the SM have been proposed to bring it to an issue \cite{peccei}. However, the status of this parameter within the SM itself is already a subject of some controversy.
In this Letter, we present a coherent way to estimate weak interaction corrections to the strong $\theta$ term. In the frame of a large-$N_c$ Chiral Perturbation Theory, we consider the physical $\eta^{(\prime)}\rightarrow \pi \pi$ amplitudes. Compared to the previous quark-gluon estimates given in \cite{ellis} and \cite{khriplo}, our hadronic approach provides a direct access to the parameter $\theta \equiv \theta_\te{QFD}+\theta_\te{QCD}$ rather than to its unphysical $\theta_\te{QFD}$ and $\theta_\te{QCD}$ components. We thus overcome phase convention issues as well as $\alpha_s$ power counting problems. Concerning this latter point, our final result given in Eq.(\ref{result})
is qualitatively compatible with the one of \cite{khriplo} given in Eq.(\ref{khriplo}) although, quantitatively, it rather agrees with the numerical result of \cite{ellis} given in Eq.(\ref{ellis}).

An important point, not addressed in this Letter, is the possibility of infinite weak corrections to $\theta$ as suggested in \cite{ellis}. This would however require the study of the $\eta^{(\prime)} \rightarrow \pi \pi$ decay amplitudes beyond the tree-level approximation considered here.


\begin{thebibliography}{99}                                                                                                
\addcontentsline{toc}{section}{References}

\bibitem{jmgHerbeumont} J.~M.~Gerard, in: 2008 European School of High-Energy Physics, 2008, p.281.

\bibitem{xingjmg}
J.~M.~Gerard and Z.~z.~Xing, Phys.\ Lett.\ B {\bf 713} (2012) 29. 

\bibitem{edm1} C.~A.~Baker, et al., Phys.\ Rev.\ Lett.\  {\bf 97} (2006) 131801.

\bibitem{edm2} M.~Pospelov and A.~Ritz, Annals Phys.\  {\bf 318} (2005) 119.

\bibitem{thooft} G.~'t Hooft, Phys.\ Rev.\ Lett.\  {\bf 37} (1976) 8.

\bibitem {ellis} J.~R.~Ellis and M.~K.~Gaillard, Nucl.\ Phys.\ B {\bf 150} (1979) 141.

\bibitem {khriplo} I.~B.~Khriplovich, Phys.\ Lett.\ B {\bf 173} (1986) 193.

\bibitem {trahern} C.~Rosenzweig, J.~Schechter and C.~G.~Trahern, Phys.\ Rev.\ D {\bf 21} (1980) 3388.

\bibitem {veneziano} P.~Di Vecchia and G.~Veneziano, Nucl.\ Phys.\ B {\bf 171} (1980) 253.

\bibitem {witten} E.~Witten, Annals Phys.\  {\bf 128} (1980) 363.

\bibitem{coleman1} S.~R.~Coleman, J.~Wess and B.~Zumino, Phys.\ Rev.\  {\bf 177} (1969) 2239.

\bibitem{coleman2} C.~G.~Callan, S.~R.~Coleman, J.~Wess and B.~Zumino, Phys.\ Rev.\  {\bf 177} (1969) 2247.

\bibitem {wittenNc} E.~Witten, Nucl.\ Phys.\ B {\bf 156} (1979) 269.

\bibitem{weinberg} S.~Weinberg, Phys.\ Rev.\ D {\bf 11} (1975) 3583.

\bibitem{georgi} H.~Georgi, Phys.\ Rev.\ D {\bf 49} (1994) 1666.

\bibitem {jmgkou} J.~M.~Gerard and E.~Kou, Phys.\ Lett.\ B {\bf 616} (2005) 85.

\bibitem {pichcp} A.~Pich and E.~de~Rafael, Nucl.\ Phys.\ B \textbf{367} (1991) 313.

\bibitem{PDG} K.~Nakamura, et al., J.\ Phys.\  G {\bf 37} (2010) 075021.

\bibitem {shaba} C.~Jarlskog and E.~Shabalin, Phys.\ Rev.\ D \textbf{52} (1995) 248.

\bibitem {jmgcs} J.~M.~Gerard, C.~Smith and S.~Trine, Nucl.\ Phys.\ B \textbf{730} (2005) 1.

\bibitem{crewther}  R.~J.~Crewther, Nucl.\ Phys.\ B {\bf 264} (1986) 277.

\bibitem{gaillardlee} M.~K.~Gaillard and B.~W.~Lee, Phys.\ Rev.\ Lett.\  {\bf 33} (1974) 108.

\bibitem{altarellimaiani} G.~Altarelli and L.~Maiani, Phys.\ Lett.\ B {\bf 52} (1974) 351.

\bibitem{vainshtein} M.~A.~Shifman, A.~I.~Vainshtein and V.~I.~Zakharov,
Sov.\ Phys.\ JETP {\bf 45} (1977) 670.

\bibitem{jmgmeson} W.~A.~Bardeen, A.~J.~Buras and J.~M.~Gerard, Phys.\ Lett.\ B {\bf 192} (1987) 138.

\bibitem{acta} J.~M.~Gerard, Acta Phys.\ Polon.\ B {\bf 21} (1990) 257.

\bibitem{jmgmeson2} A.~J.~Buras and J.~M.~Gerard, Phys.\ Lett.\ B {\bf 192} (1987) 156.

\bibitem{Buchalla} G.~Buchalla, A.~J.~Buras and M.~E.~Lautenbacher, Rev.\ Mod.\ Phys.\  {\bf 68} (1996) 1125.

\bibitem{flag} A.~Juttner, Review: The FLAG working group, in: 14th International Conference on Hadron Spectroscopy (Hadron 2011), arXiv:1109.1388 [hep-ph].

\bibitem{J} C.~Jarlskog, Phys.\ Rev.\ Lett.\  {\bf 55} (1985) 1039.

\bibitem{ciri} V.~Cirigliano, G.~Ecker, H.~Neufeld, A.~Pich and J.~Portoles, Rev.\ Mod.\ Phys.\  {\bf 84} (2012) 399.


\bibitem{jmgBur} A.~J.~Buras and J.~M.~Gerard, Phys.\ Lett.\ B {\bf 517} (2001) 129.

\bibitem{theta}  R.~J.~Crewther, P.~Di Vecchia, G.~Veneziano and E.~Witten, Phys.\ Lett.\ B {\bf 88} (1979) 123.

\bibitem{peccei} R. D. Peccei, Lect. Notes Phys. 741 (2008) 3.

\end{thebibliography}
\end{document}